\documentclass[lettersize,journal]{IEEEtran}
\usepackage{amsmath,amsfonts}
\usepackage{algorithmic}
\usepackage{algorithm}
\usepackage{array}
\usepackage[caption=false,font=normalsize,labelfont=sf,textfont=sf]{subfig}
\usepackage{textcomp}
\usepackage{stfloats}
\usepackage{url}
\usepackage{verbatim}
\usepackage{graphicx}
\usepackage{cite}
\usepackage[table,xcdraw]{xcolor}
\usepackage{threeparttable}
\begin{document}

\title{A 65 nm Trustworthy Hypoglycemia Forecasting Engine \\Achieving 11.3 nJ per Inference}

\author{Boyang Cheng,~\IEEEmembership{Student Member,~IEEE,} 
        Jianbo Liu,~\IEEEmembership{Student Member,~IEEE,} 
        Pengyu Ren,~\IEEEmembership{Student Member,~IEEE,} 
        Xueji Zhao,~\IEEEmembership{Student Member,~IEEE,} 
        Steven Davis,~\IEEEmembership{Student Member,~IEEE,} 
        Likai Pei,~\IEEEmembership{Student Member,~IEEE,} 
        Zephan M. Enciso,~\IEEEmembership{Student Member,~IEEE,} 
        Kai Ni,~\IEEEmembership{Member,~IEEE,} 
        Ningyuan Cao,~\IEEEmembership{Member,~IEEE,}
        % <-this % stops a space
\thanks{This work has been submitted to the IEEE for possible publication. Copyright may be transferred without notice, after which this version may no longer be accessible.}}
% The paper headers
\markboth{}%
{Shell \MakeLowercase{\textit{et al.}}: A Sample Article Using IEEEtran.cls for IEEE Journals}

% \IEEEpubid{0000--0000/00\$00.00~\copyright~2021 IEEE}
% Remember, if you use this you must call \IEEEpubidadjcol in the second
% column for its text to clear the IEEEpubid mark.

\maketitle

\begin{abstract}
Diabetes affect approximately 38.4 million people in the United States. Continuous glucose monitoring (CGM) devices provide real-time insight into glucose dynamics allowing people to forecast hypoglycemia events. The adoption of AI has enabled more accurate predictive modeling. However, biomedical applications require transparency and explainability, making traditional black-box AI models unsuitable for applications such as glucose monitoring. To address these challenges, this paper presents a 65 nm hypoglycemia-forecasting engine based on probabilistic decision trees (PDTs) for noise-robust, explainable medical inference. A reconfigurable 4 × 24 × 24 probabilistic-node (pNode) array enables scalable decision sampling for arbitrary tree with a maximum depth of 12, coordinated by an on-chip low-power RISC-V core. The chip achieves 11.3 nJ/inference, a state-of-the-art 30-min forecasting F1 of 0.825, and 4.1–16.1× improved robustness to noise and data-point drop-off.
\end{abstract}


\section{Introduction}
\label{sec:introduction}
\begin{figure}[!t]
\centering
\includegraphics[width=\columnwidth]{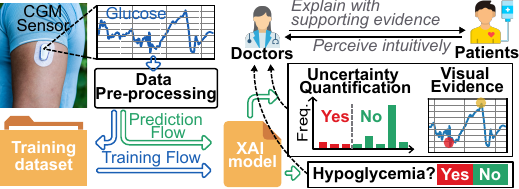}
\caption{Trustworthiness is essential for CGM analytics, requiring interpretable and uncertainty-aware decisions.}
% \vspace{-0.3cm}
\label{fig:cgm_sensor_app}
\end{figure}

\begin{figure}[!b]
\centering
\includegraphics[width=\columnwidth]{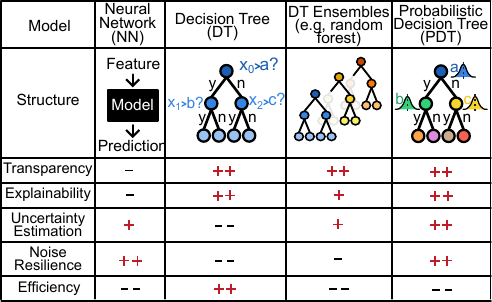}
\caption{Trustworthiness comparison across different algorithms. Probabilistic decision trees (PDTs) offer the outstanding trustworthiness but suffers from poor efficiency.}
\label{fig:algorithm_table}
\end{figure}

\IEEEPARstart{D}{iabetes} is a chronic metabolic disorder characterized by persistently elevated blood glucose levels resulting from impaired insulin secretion, reduced insulin sensitivity, or a combination of both physiological dysfunctions. In the U.S., an estimated 40.1 million people, about 12.0\% of the population, are affected \cite{noauthor_national_2026}. Continuous glucose monitoring (CGM) devices provide real-time insight into glucose dynamics, with the global market projected to reach USD 17.1 billion by 2030 \cite{noauthor_continuous_2026}. A major and potentially life-threatening complication is hypoglycemia, where blood glucose drops to dangerously low levels. Traditional threshold-based alarms often fail to provide timely warnings, motivating the use of artificial intelligence (AI) to predict low-glucose events and enable pre-emptive intervention \cite{lin_alarm_2020, xie_benchmarking_2020}. In mission-critical medical applications, AI models must satisfy strict requirements for transparency and explainability, with the World Health Organization (WHO) identifying “ensuring transparency, explainability and intelligibility of AI systems” as a core ethical principle for health-care AI \cite{noauthor_ethics_2021}, enabling clinicians to justify decisions and patients to perceive alerts intuitively, as shown by Fig.~\ref{fig:cgm_sensor_app}. Under these constraints, black-box neural networks become difficult to certify. 
To address the black-box nature of conventional models, Bayesian neural network engines were proposed for ventricular arrhythmia detection by introducing Bayesian fully connected layers into a CNN for uncertainty estimation \cite{liu_bnn_2025, cai_vibnn_2018, enciso_bnn_2026}. While symbolic AI approaches \cite{cheng_neuromorphi_2024, liu_neuro_symblic_2026} offer inherent interpretability for biomedical applications, they often come at the cost of a large memory footprint. Post-hoc approaches such as Gradient Backpropagation \cite{bhat_gradient_2022} provide saliency-like evidence but still lack transparency and hard to deployed on edge devices. Decision Trees (DTs) have long been valued as trustworthy models because they expose their full decision path, unlike black-box deep neural networks. However, conventional DTs rely on hard decision boundaries and are therefore highly sensitive to sensor noise and patient variability—both common in CGM signals—and cannot quantify uncertainty in their predictions. Ensemble methods such as random forests (RFs) \cite{seo_machine-learning_2019} can partially improve robustness but still inherit the hard-boundary behavior of individual trees and lose explainability, as their outputs reflect the aggregation of many trees rather than a single interpretable reasoning process. Probabilistic Decision Trees (PDTs) are a promising approach for transparency and explainability. Unlike conventional decision trees where each node uses a fixed threshold, Each node in PDT produces a probability-driven decision rather than a hard comparison, allowing uncertainty to propagate through the tree. Unlike black-box deep neural networks, PDTs provide explicit reasoning over decision paths while capturing uncertainty through probabilistic evaluation. Moreover, PDTs are valued for their robustness to noisy real-world data a limitation commonly observed in conventional DTs and RFs, and provide more trustworthiness through uncertainty estimation, as shown by Fig.~\ref{fig:algorithm_table}.

% \begin{figure}[!b]
% \centering
% \includegraphics[width=\columnwidth]{figure/fig2_radar_n_PDT.pdf}
% \caption{Probabilistic decision trees offer the outstanding trustworthiness but suffers from poor efficiency.}
% \label{fig:radar_n_PDT}
% \end{figure}

\begin{figure}[!t]
\centering
\includegraphics[width=\columnwidth]{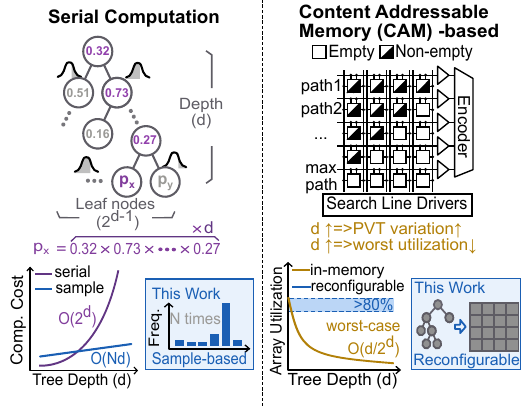}
\caption{Compute complexity for PDT on von Neumann processors. Non-von Neumann architectures such as CAM show reduced hardware utilization when operating on spare PDTs}
% \vspace{-0.3cm}
\label{fig:complexity_utilization}
\end{figure}

One major challenge in implementing deep PDTs in von Neumann processors is the exponentially increasing inference complexity of $O(2^d)$ for the depth of the tree, d. This complexity arises from both the heavy floating-point arithmetic required for probability aggregation and the irregular memory access imposed by arbitrary tree structures (Fig.~\ref{fig:complexity_utilization}, left). The additional need to compute probabilities from different distributions further increases the computational overhead. In contrast, non–Von Neumann in-memory tree search \cite{pedretti_tree-based_2021} can improve efficiency by avoiding serial computation, but suffers from hardware inefficiencies, including poor memory utilization in unbalanced trees and limited scalability in deeper paths, where accumulated device variations and analog noise significantly degrade performance (Fig.~\ref{fig:complexity_utilization}, right). To address these challenges, this paper presents a 65 nm reconfigurable hypoglycemia-forecasting engine based on PDTs for noise-robust and explainable medical inference. 

The remainder of this paper is structured as follows: Section~\ref{sec:background} reviews the background of hypoglycemia forecasting and summarizes representative prior work, followed by an introduction to the probabilistic sample-based soft decision tree methodology. Section~\ref{sec:hybrid_pdt_engine_architecture} presents the hybrid PDT engine architecture that supports both arithmetic-based and sample-based approaches for efficient PDT inference. Section~\ref{sec:design_of_pdt_engine_circuits} introduces the design of circuits at the node level (pNode) and system level. Section~\ref{sec:evaluation} shows the measured hardware and algorithm results on a prototype chip fabricated in TSMC 65 nm CMOS technology. Finally, Section~\ref{sec:conclusion} makes a conclusion.

\section{Background}
\label{sec:background}
\subsection{AI for Hypoglycemia Forecasting}
\label{sec:hypoglycemia_forecasting}
In recent years, blood glucose (BG) monitoring has been revolutionized by the advent of CGM sensors, consisting of wearable subcutaneous needle-based and minimally-invasive devices that allow measuring the BG concentration almost continuously (1–5 min sampling period) for several consecutive days/weeks. Several algorithms for the real-time prediction of hypoglycemic events from CGM data have been proposed.

In hypoglycemia forecasting, precision and sensitivity are standard classification metrics used to evaluate how reliably a model predicts hypoglycemic events (typically defined as glucose $<$ 70 mg/dL \cite{ada_glycemic_2023}). These metrics are computed from the confusion matrix: true positives (TP) are correctly predicted hypoglycemia events, false positives (FP) are normal glucose periods incorrectly predicted as hypoglycemia, and false negatives (FN) are missed hypoglycemia events. Sensitivity is defined as:
\begin{equation}
    Sensitivity = \frac{TP}{TP + FN}
\end{equation}
and measures how many real hypoglycemia events are successfully detected. In forecasting applications, high sensitivity is critical because missing a hypoglycemia event can directly impact patient safety. Precision is defined as:
\begin{equation}
    Precision = \frac{TP}{TP + FP}
\end{equation}
and measures how many predicted hypoglycemia alarms are actually correct. High precision reduces false alarms, which is important for patient trust, and alarm fatigue. In practical CGM hypoglycemia forecasting, there is often a trade-off between sensitivity and precision: increasing sensitivity (catching more true events) may increase false alarms, while increasing precision may risk missing some events. The F1 score is a harmonic-mean metric that combines precision and sensitivity into a single value, providing a balanced measure of classification performance when both false alarms and missed hypoglycemia events are important. It is computed as:
\begin{equation}
    F1 = 2 \cdot \frac{Precision \cdot Sensitivity }{ Precision + Sensitivity}
\end{equation}

Prior work on CGM-based hypoglycemia forecasting AI can be broadly organized into (i) rule-based models, (ii) classic machine learning models, such as Random Forests (RFs) and Support Vector Machines (SVMs), and (iii) Deep temporal models, such as Recurrent Neural Networks (RNNs) and Long Short-Term Memories (LSTMs). In the first category, M. De La Cruz et al. \cite{de_la_cruz_explainable_2024} proposed a rule-based interpretable hypoglycemia prediction model based on Dynamic Structured Grammatical Evolution, achieving accurate hypoglycemia forecasting while maintaining human-readable rule-level clinical interpretability. In the second category, W. Seo et al. \cite{seo_machine-learning_2019} demonstrated 30-min prediction horizon random forest achieved strong discrimination with high sensitivity/specificity, highlighting the practicality of ensemble classifiers for near-term warning generation. M. Gadaleta et al. \cite{gadaleta_prediction_2019} proposed a SVM-based model covering both regression and classification formulations as well as static vs. dynamic training strategies. In the third category, T. Zhu et al. \cite{zhu_personalized_2023} proposed an attention-based RNN with an evidential output layer to provide theoretically grounded prediction confidence, and they further used model-agnostic meta-learning to adapt quickly to new patients. M. Yang et al. \cite{yang_joint_2022} proposed a LSTM-based deep multi-task learning for glucose forecasting and hypoglycemia event prediction and reported the classification branch significantly outperforms the forecasting branch on hypoglycemia prediction. However, limited research has investigated how these algorithms perform when implemented in hardware. This work introduces a PDT based hypoglycemia forecasting engine which considers the trade-offs among algorithmic complexity, performance, and trustworthiness, and is validated through post-silicon evaluation, as described in the following sections.

\subsection{Probabilistic Decision Tree}
A hard decision tree is an directed acyclic graph. All its nodes have a parent node except the root node, the only one that has no parent. The remaining nodes can be categorized into two types: intermediate nodes and leaf nodes. Intermediate nodes implement decision rules based on feature thresholds or Boolean conditions. Each intermediate node evaluates a specific feature (or a function of features) and routes the input to one of its child branches according to the comparison outcome. Leaf nodes, in contrast, terminate the decision path. They do not perform further splitting but instead produce the final prediction result. The inference process of a hard decision tree can be viewed as a deterministic traversal from the root to a single leaf node, where each intermediate decision progressively refines the partition of the input space until a terminal prediction is reached. Soft decision trees replace deterministic branching with probabilistic routing. Instead of sending an input sample exclusively to a single child node, each intermediate node computes a routing probability that softly distributes the sample to multiple branches \cite{frosst_soft_2017}. As a result, the inference process becomes a weighted aggregation over multiple root-to-leaf paths rather than a single deterministic path.

Formally, in a hard decision tree, the decision function at an intermediate node $i$ can be written as a binary indicator:
\begin{equation}
d_i(\mathbf{x}) =
\begin{cases}
1, & \text{if } \mathbf{w}_i^\top \mathbf{x} - b_i \ge 0 \\
0, & \text{otherwise}
\end{cases}
\end{equation}
where $\mathbf{w}_i$ is the weights, $\mathbf{x}$ is the feature, and $b_i$ is the threshold. The routing is discrete: if $d_i(\mathbf{x}) =1$, route to the left child; otherwise, route to the right child. Consequently, the final output corresponds $\hat{y}(\mathbf{x})$ to a single leaf node $\ell$:
\begin{equation}
\hat{y}(\mathbf{x}) 
= \sum_{\ell \in \mathcal{L}} 
\left(
\prod_{i \in \text{path}(\ell)}
\mathbf{1}\big( d_i(\mathbf{x}) = \alpha_{i,\ell} \big)
\right)
y_\ell
\end{equation}
where $\mathcal{L}$ denotes the set of all leaf nodes, $\alpha_{i,\ell} \in \{0, 1\}$ is the direction of the path, and $y_\ell$ is the output of each leaf node. In a soft decision tree, the hard indicator function is replaced by a gating function $\sigma(\cdot)$:
\begin{equation}
p_i(\mathbf{x}) = \sigma\big(\mathbf{w}_i^\top \mathbf{x} - b_i\big)
\end{equation}
where $p_i(\mathbf{x})$ represents the probability of routing the sample to one child, while $1-p_i(\mathbf{x})$ corresponds to the other child. The probability of reaching a leaf node $\ell$ is then given by the product of routing probabilities along the path:

\begin{equation}
P_\ell(\mathbf{x}) = \prod_{i \in \text{path}(\ell)} 
\big[p_i(\mathbf{x})\big]^{\alpha_{i,\ell}}
\big[1 - p_i(\mathbf{x})\big]^{1-\alpha_{i,\ell}}
\end{equation}
and the final prediction becomes a weighted aggregation over all leaves:
\begin{equation}
\hat{y}(\mathbf{x}) = \sum_{\ell} P_\ell(\mathbf{x}) \, y_\ell
\end{equation}
For a binary soft decision tree with depth $d$ having an exponential time complexity of $\mathcal{O}(2^d)$ per inference. To avoid the exponential cost of exact soft aggregation, we approximate the soft decision tree output via probabilistic sampling by interpreting the soft routing as a stochastic path-selection process. Specifically, at each intermediate node $i$, we sample a binary routing decision $z_i \in \{0,1\}$ according to the gating probability $p_i(\mathbf{x})$"

\begin{equation}
z_i \sim \mathrm{Bernoulli}\!\left(p_i(\mathbf{x})\right),
\end{equation}

where $z_i=1$ selects one child (e.g., left) and $z_i=0$ selects the other (e.g., right). A single sampled sequence $\mathbf{z}=(z_1,\dots,z_D)$ uniquely determines a sampled leaf $\ell(\mathbf{z})$ and yields a single-path prediction $y_{\ell(\mathbf{z})}$. Under this stochastic traversal, the soft decision tree output can be written as an expectation over the induced leaf distribution:

\begin{equation}
\hat{y}(\mathbf{x}) = \sum_{\ell} P_\ell(\mathbf{x})\, y_\ell
= \mathbb{E}_{\ell \sim P(\cdot \mid \mathbf{x})}\!\left[y_\ell\right].
\end{equation}

We then approximate this expectation with $N$ independent samples:

\begin{equation}
\hat{y}_{\text{prob.}}(\mathbf{x})
= \frac{1}{N}\sum_{n=1}^{N} y_{\ell^{(n)}}
\end{equation}
where each $\ell^{(n)}$ is obtained by one stochastic root-to-leaf traversal. By construction, $\hat{y}_{\text{prob.}}(\mathbf{x})$ is an unbiased estimator of the exact soft output:

\begin{equation}
\mathbb{E}\!\left[\hat{y}_{\text{prob.}}(\mathbf{x})\right]
= \hat{y}(\mathbf{x}),
\end{equation}
reducing the computational complexity from $\mathcal{O}(2^d)$ to $\mathcal{O}(Nd)$. In the probabilistic approximation, the independent sample count governs the accuracy-complexity trade off. Since $\hat{y}_{\text{prob.}}$ is an estimator of the exact output $\hat{y}$: 
\begin{equation}
\label{eq:estimator}
\hat{y}_{\text{prob.}}(\mathbf{x}) 
= \frac{1}{N}\sum_{n=1}^{N} y_{\ell^{(n)}}
\end{equation}
its estimation error is characterized by its variance:
\begin{equation}
\mathbb{E}\left[\hat{y}_{\text{prob.}}(\mathbf{x})\right]=\hat{y}(\mathbf{x}), \qquad
\mathrm{Var}\left[\hat{y}_{\text{prob.}}(\mathbf{x})\right]
= \frac{1}{N}\mathrm{Var}\left[y_{\ell}\right]
\end{equation}
And this implies a $\mathcal{O}(1/\sqrt{N})$ convergence rate in root-mean-squared error. Based on Hoeffding's inequality

\begin{equation}
\delta = \mathbb{P}\left(\|\hat{y}_{\text{prob.}}(\mathbf{x}) - \hat{y}(\mathbf{x})\|_2 \ge \varepsilon \right) \le 2\exp\left(
-2N\varepsilon^2\right)
\end{equation}
to achieve the target error tolerance $\varepsilon$, and confidence level 1-$\delta$:
\begin{equation}
N \ge \frac{1}{2\varepsilon^2} \ln\left(\frac{2}{\delta}\right)
\end{equation}

\section{Hybrid PDT Engine Architecture}
\label{sec:hybrid_pdt_engine_architecture}

\begin{figure}[!t]
\centering
\includegraphics[width=\columnwidth]{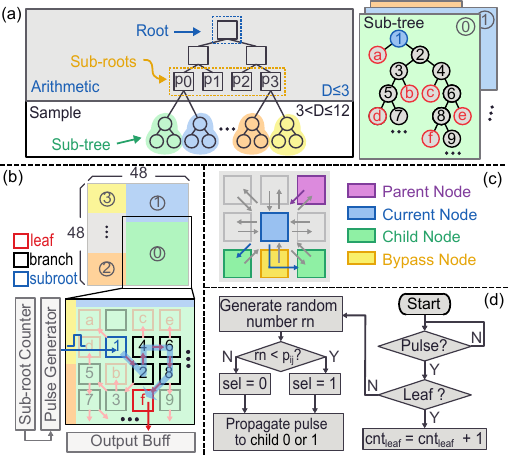}
\caption{(a)~Hybrid PDT engine. (b)~Mapping the sub-trees onto the array. (c)~Reconfigurable 8-way bidirectional pNode with bypass functionality. (d)~Operating flowchart of pNode.}
% \vspace{-0.3cm}
\label{fig:hybrid_structure}
\end{figure}

\begin{figure}[!b]
\centering
\includegraphics[width=0.8\columnwidth]{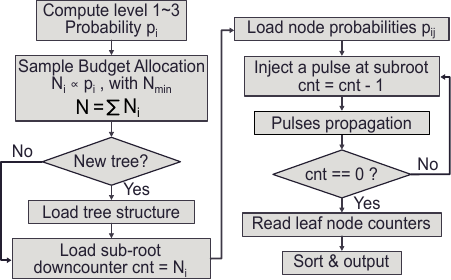}
\caption{Operating flow for hybrid probabilistic decision tree engine.}
\label{fig:flowchart}
\end{figure}

The PDT inference engine features a hybrid of computation and sampling methods, as shown in Fig.~\ref{fig:hybrid_structure}~(a). For shallow layers ($D \leq 3$), leaf-node probabilities are computed exactly by a statistical solver. For deep layers ($3 < D \leq 12$), remaining sub-trees are mapped onto a 4$\times$24$\times$24 reconfigurable pNode array for sampling-based inference, as shown by Fig.~\ref{fig:hybrid_structure}~(b). The sub-tree–array mapping is performed in two phases, Configuration and Sampling. In the Configuration phase, each pNode is programmed with structural information, including its role as a sub-root, branch, leaf, or bypass node, as well as the indices of its two child nodes. The in-node register files are externally accessible, enabling the chip to be quickly reprogrammed with different trees or parameters. To accommodate placement constraints where direct parent–child connections are unavailable, a pNode can operate in bypass mode with a single input and output. Together with 8-way bidirectional connectivity, this bypass capability greatly enhances flexibility in mapping arbitrary tree structures while maintaining high array utilization, as shown by Fig.~\ref{fig:hybrid_structure}~(c). 

In the Sampling phase, inference is performed by propagating pulses through pNode connections, as shown by Fig.~\ref{fig:hybrid_structure}~(d). Each pNode supports probabilistic branching: when activated by an input pulse from its parent or from a pulse generator, the cell generates a random number, compares it with its stored probability, and forwards the pulse to one of its child nodes accordingly. The branching probabilities are pre-loaded into each pNode register before sampling begins. Each sub-root node is assigned a down-counter and a sampling budget Ni, proportional to the probability computed by the statistical solver. A pulse generator injects Ni pulses into the sub-root nodes, which then propagate through the pNode array from parent to child until reaching the leaves. At leaf nodes, counters accumulate the number of arriving pulses, producing probability estimates without requiring floating-point arithmetic. Once a down-counter reaches zero, indicating that sampling of the sub-tree is complete, the total number of pulses collected at each leaf represents the inference results, which can then be ranked or sorted depending on the application, as shown by Fig.~\ref{fig:flowchart}. This architecture eliminates the need for centralized arithmetic units and instead takes advantage of distributed pulse-based computation, achieving both low inference complexity of $O(Nd)$, where N is the total sampling budget, and high mapping efficiency.

\section{Design of PDT Engine Circuits}
\label{sec:design_of_pdt_engine_circuits}

\begin{figure}[!t]
\centering
\includegraphics[width=\columnwidth]{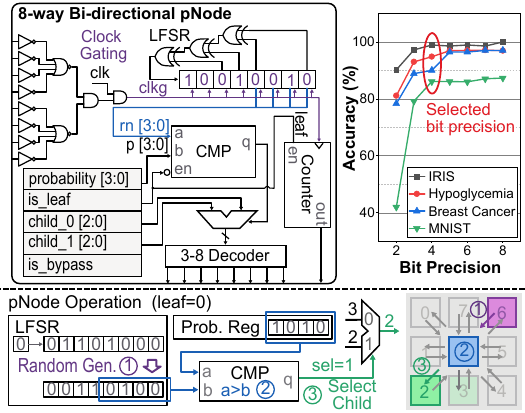}
\caption{pNode circuit architecture and its operation when configured as an intermediate node. Experiments on different datasets show that a 4-bit system has the best trade-off between accuracy and hardware overhead.}
% \vspace{-0.3cm}
\label{fig:pNode_structure}
\end{figure}

\begin{figure}[!t]
\centering
\includegraphics[width=\columnwidth]{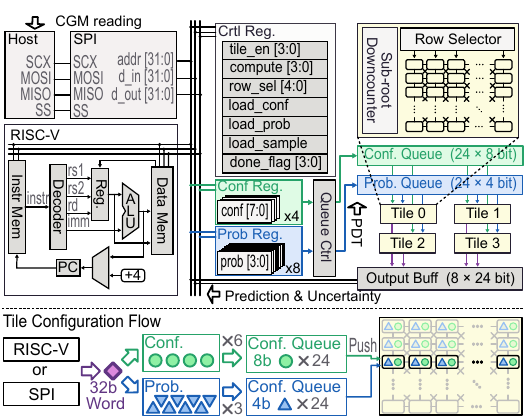}
\caption{Hybrid PDT system architecture. It consists of four 24×24 pNode tiles, totaling 2,304 pNodes. The configuration and probability queues are separated to optimize the data flow and facilitate model fine-tuning.}
% \vspace{-0.3cm}
\label{fig:system}
\end{figure}

Fig.~\ref{fig:pNode_structure} shows the design of a pNode. Each pNode is composed of an input stage, core logic, and an output stage. At the fan-in stage, up to 8 fan-in connections and incorporates local clock gating to disable inactive paths, thereby reducing dynamic power when pulses are absent. The core logic integrates a small register file, a linear feedback shift register (LFSR), and a comparator. Algorithmic validation results show that when the bit precision of probability exceeds 4 bits and the length of LFSR exceeds 8 bits, the improvement in accuracy becomes marginal. To balance hardware cost and algorithmic performance, the bit precision of random numbers in this work is therefore set to 4. To mitigate cyclic artifacts in the LFSR sequence and to match the probability bit precision, only 4 bits are extracted as the effective random value, aligned directly with the 4-bit probability register. The output stage selects one child node based on the comparison between the LFSR (rn) and the probability value (p) and forwards a pulse to it. When a pNode is configured as a leaf node, the LFSR and comparator are disabled, and an 8-bit counter is enabled to accumulate the arriving pulses, generating class probability distributions directly in hardware, thereby avoiding unnecessary switching activity and improving energy efficiency.

\begin{figure}[!b]
\centering
\includegraphics[width=0.9\columnwidth]{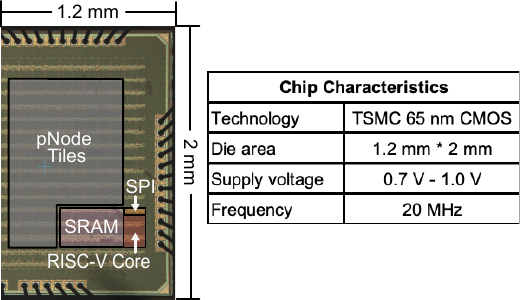}
\caption{Photo of a 65 nm prototype chip and characteristics.}
\label{fig:die_photo}
\end{figure}

As shown by Fig.~\ref{fig:system}, the entire hypoglycemia forecasting engine integrates four 24~×~24 pNode tiles, a RV32I RISC-V core, peripheral configuration logic, and on-chip register files for control and data management. Each tile is a 24~×~24 array of pNodes while the RISC-V core provides supervisory functions such as configuration sequencing, sorting, and probability normalization. Control and configuration registers manage tile activation, row addressing, and data flow between the processor, SPI interface, and the tiles. Together, these components form a heterogeneous system that balances high-throughput stochastic computing in the arrays with flexible digital control. These 4 tiles support dynamically reconfiguration to support parallel sampling. When the sampling budget of a particular sub-tree is high, other tiles can be reloaded to perform sampling concurrently, thereby improving throughput and balancing computation across the array.

To optimize the configuration flow of pNode tiles, configuration data and probability parameters are delivered separately through 32-bit words from either the RISC-V processor or the SPI interface. To match the much wider row requirements of the tiles, dedicated queues are introduced as interface bridges. A full configuration row requires 192 bits (24 nodes × 8 bits), and a probability row requires 96 bits (24 nodes × 4 bits). Accordingly, the configuration queue aggregates six 32-bit words, and the probability queue aggregates three words, before committing a complete row. This batching strategy aligns the external interface with the internal data path, simplifies timing and reduces control overhead. An example timing diagram is presented in Fig. 4 to illustrate the operating sequence. Once accumulation within a tile completes, a done flag is raised in control register. The RISC-V core can detect this through polling or interrupts, then retrieve counter values in each leaf pNode, perform sorting or normalization, and reconfigure the tile for the next task. This handshake bridges massively parallel inference in the arrays with supervisory control in the processor. 

\section{Evaluation}
\label{sec:evaluation}

\begin{figure}[!t]
\centering
\includegraphics[width=0.9\columnwidth]{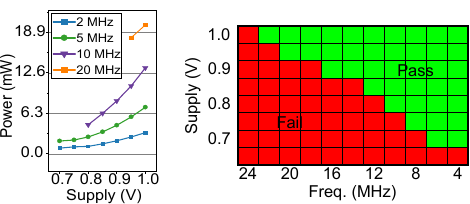}
\caption{Measured dynamic power and shmoo plot.}
\label{fig:power_shmoo}
\end{figure}

\begin{figure}[!t]
\centering
\includegraphics[width=\columnwidth]{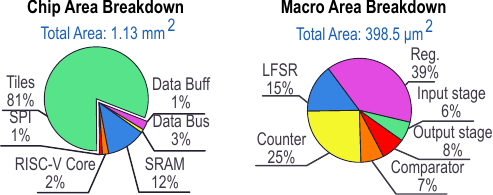}
\caption{Area breakdown of chip and pNode macro.}
% \vspace{-0.3cm}
\label{fig:area_breakdown}
\end{figure}

The prototype chip is fabricated in TSMC 65 nm CMOS technology. The die photo and its characteristics are shown in Fig.~\ref{fig:die_photo}. The chip operates over a supply voltage range of 0.7–1.0 V and a clock frequency range of 2–22 MHz. Power measurements across 0.7–1.0 V and 2–20 MHz show 0.2 mW consumption at 0.7 V and 2 MHz, as shown by Fig.~\ref{fig:power_shmoo}. 
The area breakdown of chip and pNode macro is shown in Fig.~\ref{fig:area_breakdown}. The total area of the engine is 1.13 mm$^2$, the RISC-V core and its associate SRAM occupy 14\% of the area. The area of a single pNode macro is 398.5 $\mu$m$^2$ which achieves an area efficiency of approximately 99.6 $\mu$m$^2$/bit.

\begin{figure}[!t]
\centering
\includegraphics[width=0.8\columnwidth]{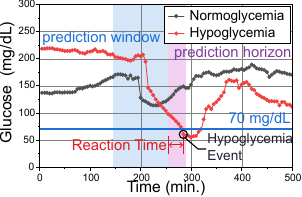}
\caption{CGM signals and example of a 2-hour window 30-min prediction horizon hypoglycemia forecasting. Glucose $<$ 70 mg/dL is defined as hypoglycemia.}
\label{fig:hypoglycemia_prediction}
\end{figure}

To evaluate the performance on hypoglycemia forecasting applications, this work is validated using the OhioT1DM CGM dataset \cite{marling_ohiot1dm_2020}. The OhioT1DM dataset is a time-stamped longitudinal dataset consisting of 5-minute-resolution CGM glucose values collected from 12 type 1 diabetes subjects over $\sim$ 8 weeks for glucose prediction and decision-support modeling. The evaluation is conducted using a 120-min prediction window and a 30-min prediction horizon, as shown in Fig.~\ref{fig:hypoglycemia_prediction}. A total of 3,000 windows were extracted and split 80/20 for training and validation. The input vector to the engine consists of 24 CGM glucose measurements along with their first- and second-order derivatives. Glucose levels below 70 mg/dL are labeled as hypoglycemia events.

\begin{table}[!t]
\caption{State-of-the-art Hypoglycemia Prediction Accuracy\label{tab:hypoglycemia_prediction_accuracy}}
\centering

\begin{threeparttable}
\begin{tabular}{|c|c|c|c|c|}
\hline
\multicolumn{1}{|c|}{\textbf{}}   & \textbf{Algorithm} & \textbf{Sensitivity$^1$} & \textbf{Precision$^2$} & \textbf{F1 Score$^3$} \\ \hline
\textbf{This Work}                & PDT                & 79.7 \%              & 85.5 \%            & 0.825             \\ \hline
\textbf{\cite{seo_machine-learning_2019}} & RF                 & 89.6 \%              & 38.9 \%            & 0.542             \\ \hline
\textbf{\cite{gadaleta_prediction_2019}} & SVM                & 86.0 \%              & 36.0 \%            & 0.508             \\ \hline
\textbf{\cite{zhu_personalized_2023}} & RNN                & 84.1 \%              & 65.6 \%            & 0.737             \\ \hline
\textbf{\cite{yang_joint_2022}} & LSTM               & 92.6 \%              & N/A                & N/A               \\ \hline
\end{tabular}

\begin{tablenotes}

\item[1] Sensitivity = True Positive / (True Positive + False Negative)
\item[2] Precision = True Positive / (True Positive + False Positive)
\item[3] F1 Score = 2 × (Precision × Sensitivity) / (Precision + Sensitivity)
\end{tablenotes}

\end{threeparttable}
% \vspace{-0.3cm}
\end{table}

\begin{figure}[!t]
\centering
\includegraphics[width=0.65\columnwidth]{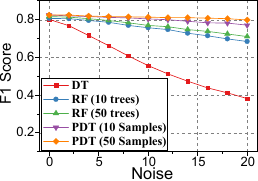}
\caption{Robustness evaluation results under different Gaussian noise levels.}
\label{fig:accuracy_vs_noise}
\end{figure}

Table \ref{tab:hypoglycemia_prediction_accuracy} presents a comprehensive comparison of event-based hypoglycemia prediction performance against state-of-the-art models. This work demonstrates consistently high precision, indicating strong capability in suppressing false hypoglycemia alarms, which is critical for improving user trust and reducing alarm fatigue in real-world deployment. Meanwhile, the achieved sensitivity remains competitive with leading SOTA approaches, ensuring reliable detection of true hypoglycemic events. By jointly optimizing precision and sensitivity, this work reflects a high F1 score highlighting its balanced and robust performance for hypoglycemia forecasting applications.

\begin{figure}[!t]
\centering
\includegraphics[width=0.65\columnwidth]{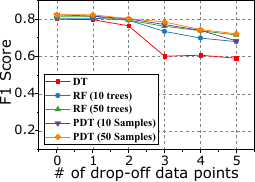}
\caption{Robustness evaluation results under different random data point drop-off.}
% \vspace{-0.3cm}
\label{fig:uncertainty}
\end{figure}

To test the  robustness, system is evaluated using two representative perturbation models that reflect real-world CGM deployment conditions. First, we inject additive absolute Gaussian noise into the glucose signal to emulate electrochemical sensing noise and physiological microenvironment variability observed in enzymatic CGM measurements. As shown in Fig.~\ref{fig:accuracy_vs_noise}, absolute Gaussian input noise with increasing intensity ($\alpha$ = 0, 2, 4, …, 20 mg/dL) is injected into the input sequence. Across all noise levels, PDT consistently achieves higher F1 scores and demonstrates 4.1–16.1× greater noise resilience ($\Delta F1 / \Delta \text{noise}$) than conventional DT and RF models. This improved robustness is attributed to the stochastic inference nature of PDT, which allows probabilistic decision aggregation rather than deterministic threshold-based decisions. Increasing the number of PDT sampling cycles or RF ensemble size further improves noise resilience by averaging out noise-induced decision variance.

Second, random data point drop-off is introduced to model real-world data incompleteness in CGM systems, including wireless packet loss during transmission, degradation at the sensor–tissue interface, and firmware-level artifact rejection or filtering. The drop-off is applied independently to both CGM glucose samples and their corresponding first- and second-order derivatives to emulate realistic signal-processing pipelines. The drop-off ratio is uniformly randomized across the input window to simulate random missing measurements. This perturbation model evaluates the system’s tolerance to incomplete input features, which is critical for continuous wearable and implantable sensing scenarios.

% absolute Gaussian input noise of increasing intensity ($\alpha$ = 0, 2, 4, …, 20 mg/dL) was added to the input data.  At the same time,  During these comparisons, number of tree nodes are constrained to be the same for DT, RF and PDT.

\begin{figure}[!t]
\centering
\includegraphics[width=0.6\columnwidth]{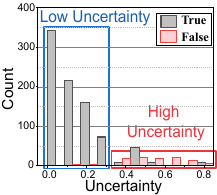}
\caption{Uncertainty estimation of PDT prediction results.}
% \vspace{-0.3cm}
\label{fig:uncertainty}
\end{figure}

We also perform uncertainty awareness evaluation shown in Fig.~\ref{fig:uncertainty}. We define uncertainty as the ratio of false predictions to true predictions. When the uncertainty is close to 0, it indicates that the model is confident in its outputs, whereas an uncertainty close to 1 suggests that the model lacks confidence in its predictions. It can be observed that most false predictions exhibit high uncertainty, whereas true predictions generally show low uncertainty. This provides an additional basis for decision-making: for example, when the model yields a high uncertainty value, it can alert the patient to the potential onset of a hypoglycemic event and prompt them to take preventive actions in advance, thereby greatly reducing the risk associated with severe hypoglycemia. Meanwhile, these high-uncertainty samples can be recorded and, together with the PDT decision paths, used to fine-tune the model to further improve accuracy or to serve as supporting evidence for medical research.

\begin{table}[!t]
\caption{Energy and Utilization Comparison\label{tab:energy_utilization}}
\centering

\begin{threeparttable}
\begin{tabular}{|lcc|lcc|}
\hline
\multicolumn{3}{|c|}{\textbf{Energy$^1$ (nJ)}}                                         & \multicolumn{3}{c|}{\textbf{Utilization$^2$ (\%)}}                                 \\ \hline
\multicolumn{1}{|l|}{\textbf{Task}}         & \multicolumn{1}{c|}{HP}     & BTSC   & \multicolumn{1}{l|}{\textbf{Task}}      & \multicolumn{1}{c|}{HP}     & BTSC   \\ \hline
\multicolumn{1}{|l|}{\textbf{This Work}}    & \multicolumn{1}{c|}{11.3}   & 6.25   & \multicolumn{1}{l|}{\textbf{This Work}} & \multicolumn{1}{c|}{96.3}   & 98.7   \\ \hline
\multicolumn{1}{|l|}{\textbf{CPU Baseline}} & \multicolumn{1}{c|}{183.3}  & 78.6   & \multicolumn{1}{l|}{\textbf{CAM}}       & \multicolumn{1}{c|}{63.4}   & 57.2   \\ \hline
\multicolumn{1}{|l|}{\textbf{Gain}}         & \multicolumn{1}{c|}{16.2×} & 12.6× & \multicolumn{1}{l|}{\textbf{Gain}}      & \multicolumn{1}{c|}{1.47×} & 1.73× \\ \hline
\end{tabular}

\begin{tablenotes}
\footnotesize
\item[1] Obtained from 65 nm simulations, considering only the dynamic energy of the multiply–accumulate operations and data access.
\item[2] Utilization = \# of nodes assigned data / \# of nodes.
\end{tablenotes}

\end{threeparttable}
% \vspace{-0.3cm}
\end{table}

To benchmark against conventional floating-point PDT implementations and CAM-based PDT architectures, which suffer from high energy consumption and limited hardware utilization, respectively, the comparison results are summarized in Table \ref{tab:energy_utilization}. The CPU baseline energy in this work is derived from 65 nm simulations of computation units and SRAM, accounting only for the dynamic energy of MAC operations and data access. Two tree configurations, hypoglycemia prediction (HP) and traffic sign recognition (BTSC) \cite{timofte_sparse_2011}, are evaluated. Leveraging the proposed hybrid compute–sample architecture, the results demonstrate 12.6× to 16.2× energy reduction compared with the CPU-based baseline. And leveraging the proposed fully reconfigurable pNode array, this work achieves 1.47× and 1.73× higher hardware utilization than CAM-based designs.

\begin{figure}[!t]
\centering
\includegraphics[width=\columnwidth]{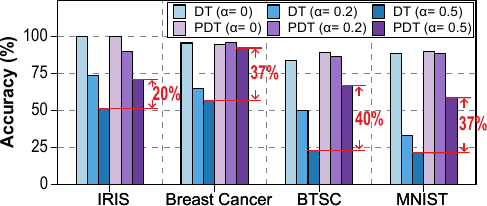}
\caption{Robustness evaluation results on other common decision tree datasets under different absolute Gaussian noise levels.}
\label{fig:accuracy_vs_normal_noise}
\end{figure}

The engine is applicable not only to hypoglycemia prediction but also to other tree-based applications, including iris classification \cite{r_a_fisher_iris_1936}, breast cancer diagnosis \cite{william_wolberg_breast_1993}, traffic sign recognition\cite{timofte_sparse_2011}, and image classification \cite{lecun_mnist_nodate}. Similarly, this work is evaluated under different noise conditions and bit-error rates, which are commonly observed in digital edge devices. 
Fig.~\ref{fig:accuracy_vs_normal_noise} shows that, with the same number of tree nodes, this work achieves 20\%–40\% higher accuracy than DT at different noise levels, and achieves 8\%-15\% higher accuracy than DT at different bit error rates.

\begin{figure}[!t]
\centering
\includegraphics[width=\columnwidth]{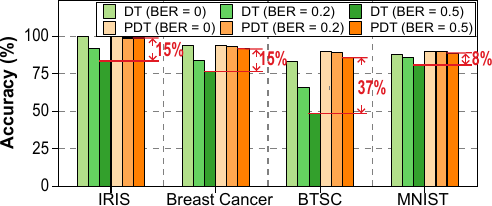}
\caption{Robustness evaluation results on other common decision tree datasets under different bit error rates (BERs).}
% \vspace{-0.3cm}
\label{fig:accuracy_vs_ber}
\end{figure}

\begin{table*}[!t]
\caption{State-of-the-art Comparison\label{tab:sota_comparison}}
\centering
\begin{threeparttable}

\begin{tabular}{|l|c|c|c|c|c|c|}
\hline
                                                                 & \textbf{This Work}                                                                                                  & \textbf{\begin{tabular}[c]{@{}c@{}}JSSC\\ 2015 \cite{lee_vocabulary_2015}\end{tabular}}          & \textbf{\begin{tabular}[c]{@{}c@{}}JSSC \\ 2018 \cite{kang_194-njdecision_2018}\end{tabular}} & \textbf{\begin{tabular}[c]{@{}c@{}}Nature Comm.\\ 2021 \cite{pedretti_tree-based_2021}\end{tabular}} & \textbf{\begin{tabular}[c]{@{}c@{}}JSSC\\ 2022 \cite{shin_neuraltree_2022}\end{tabular}}            & \textbf{\begin{tabular}[c]{@{}c@{}}TCAS-I\\ 2023 \cite{shih_dr_2023}\end{tabular}}                           \\ \hline
Process (nm)                                                     & 65                                                                                                                  & 65                                                                    & 65                                                            & 65                                                                   & 65                                                                      & 40                                                                                       \\ \hline
Technology                                                       & Digital CMOS                                                                                                        & Digital CMOS                                                          & CIM                                                           & ACAM (sim.)                                                          & Digital CMOS                                                            & Digital CMOS (sim.)                                                                      \\ \hline
Algorithm                                                        & PDT                                                                                                                 & RF                                                                    & DT                                                            & DT                                                                   & NeuralTree                                                              & DT                                                                                       \\ \hline
Freq. (MHz)                                                      & 20                                                                                                                  & 250                                                                   & 1000 (Ctrl Only)                                              & 1                                                                    & --                                                                      & 429                                                                                      \\ \hline
Supply (V)                                                       & 0.7 - 1.0                                                                                                           & 1.2                                                                   & 1.0                                                           & 1.0                                                                  & 1.2                                                                     & 0.9                                                                                      \\ \hline
Area (mm$^2$)                                                       & 1.13                                                                                                                & 2.3                                                                   & 1                                                             & --                                                                   & 8                                                                       & 2.12                                                                                     \\ \hline
Area/bit (um$^2$)                                                   &  \textbf{99.6}                                                                                & --                                                                    & --                                                            & 120                                                                  & --                                                                      & --                                                                                       \\ \hline
\# of Nodes                                                      &  \textbf{2304}                                                                                & --                                                                    & --                                                            & 2304                                                                 & --                                                                      & --                                                                                       \\ \hline
Max Tree Depth                                                   &  \textbf{12}                                                                                  & --                                                                    & 6                                                             & 10                                                                   & 4                                                                       & 8                                                                                        \\ \hline
Energy/sample. (nJ)                                              &  \textbf{0.226$^*$}                                                                               & 93.5                                                                  & 19.4                                                          & 1.28                                                                 & 227                                                                     & 0.172                                                                                    \\ \hline
\begin{tabular}[c]{@{}l@{}}Uncertainty \\ Awareness\end{tabular} &  \textbf{Yes}                                                                                 & No                                                                    & No                                                            & No                                                                   & No                                                                      & No                                                                                       \\ \hline
Noise Robustness                                                 &  \textbf{Yes}                                                                                 & Yes                                                                   & No                                                            & No                                                                   & Yes                                                                     & No                                                                                       \\ \hline
\begin{tabular}[c]{@{}l@{}}Task \&\\ Accuracy (\%)\end{tabular}  & \begin{tabular}[c]{@{}c@{}}Hypoglycemia Prediction \\ (96.9)\\ BTSC \\ (90.3)\\ Breast Cancer\\ (96.5)\end{tabular} & \begin{tabular}[c]{@{}c@{}}Object\\ Recognition\\ (93.5)\end{tabular} & \begin{tabular}[c]{@{}c@{}}BTSC\\ (93)\end{tabular}           & \begin{tabular}[c]{@{}c@{}}BTSC\\ (96.5)\end{tabular}                & \begin{tabular}[c]{@{}c@{}}Seizure \\ Detection\\ (95.6**)\end{tabular} & \begin{tabular}[c]{@{}c@{}}Breast Cancer Coimbra\\ (72.4)\\ Stroke\\ (93.7)\end{tabular} \\ \hline
\end{tabular}

\begin{tablenotes}

\item[*] Energy is measured at 0.7 V, 2 MHz operating frequency, 231 tree nodes.

\end{tablenotes}

\end{threeparttable}
\end{table*}

\section{Conclusion}
\label{sec:conclusion}

This work presents a 65 nm hypoglycemia-forecasting engine based on probabilistic decision trees (PDTs) for noise-robust, explainable medical inference. A sampling-based approach is employed to approximate the soft decision tree in an energy-efficient manner, enabling its integration into energy-constrained bio-medical devices. A reconfigurable 4~×~24~×~24 probabilistic-node array enables scalable decision sampling for arbitrary tree structures, coordinated by an on-chip low-power RISC-V MCU. As shown by Table \ref{tab:sota_comparison}, this engine achieves 11.3 nJ/inference, a state-of-the-art 30-min forecasting F1 of 0.825, and 4.1–16.1× improved robustness to CGM sensor noise compared to prior designs. 

% \newpage
\vspace{0.5cm}

\bibliographystyle{IEEEtran}
%\bibliographystyle{named}
% \bibliography{reference}
\bibliography{TCAS-I_probabilistic_decision_tree}

@misc{lecun_mnist_nodate,
	title = {{MNIST} handwritten digit database},
	url = {http://yann.lecun.com/exdb/mnist},
	author = {LeCun, Yann and Cortes, Corinna},
}

@techreport{noauthor_national_2026,
	title = {National {Diabetes} {Statistics} {Report} (2026)},
	url = {https://gis.cdc.gov/grasp/diabetes/diabetesatlas-statsreport.html},
	institution = {Centers for Disease Control and Prevention},
	month = jan,
	year = {2026},
}

@techreport{noauthor_continuous_2026,
	title = {Continuous {Glucose} {Monitoring} ({CGM}) {Market} {Size} \& {Share} {Analysis} - {Growth} {Trends} and {Forecast} (2026 - 2031).},
	url = {https://www.mordorintelligence.com/industry-reports/continuous-glucose-monitoring-market},
	institution = {Mordor Intelligence Research \& Advisory},
	month = feb,
	year = {2026},
}

@article{lin_alarm_2020,
	title = {Alarm {Settings} of {Continuous} {Glucose} {Monitoring} {Systems} and {Associations} to {Glucose} {Outcomes} in {Type} 1 {Diabetes}},
	volume = {4},
	copyright = {http://creativecommons.org/licenses/by-nc-nd/4.0/},
	issn = {2472-1972},
	doi = {10.1210/jendso/bvz005},
	language = {en},
	number = {1},
	urldate = {2026-02-17},
	journal = {Journal of the Endocrine Society},
	author = {Lin, Yu Kuei and Groat, Danielle and Chan, Owen and Hung, Man and Sharma, Anu and Varner, Michael W and Gouripeddi, Ramkiran and Facelli, Julio C and Fisher, Simon J},
	month = jan,
	year = {2020},
	pages = {bvz005},
}

@book{noauthor_ethics_2021,
	address = {Geneva},
	edition = {1st ed},
	title = {Ethics and {Governance} of {Artificial} {Intelligence} for {Health}: {WHO} {Guidance}. {Executive} {Summary}},
	isbn = {978-92-4-003740-3},
	shorttitle = {Ethics and {Governance} of {Artificial} {Intelligence} for {Health}},
	language = {eng},
	publisher = {World Health Organization},
	year = {2021},
}

@article{seo_machine-learning_2019,
	title = {A machine-learning approach to predict postprandial hypoglycemia},
	volume = {19},
	issn = {1472-6947},
	doi = {10.1186/s12911-019-0943-4},
	language = {en},
	number = {1},
	urldate = {2026-02-17},
	journal = {BMC Medical Informatics and Decision Making},
	author = {Seo, Wonju and Lee, You-Bin and Lee, Seunghyun and Jin, Sang-Man and Park, Sung-Min},
	month = dec,
	year = {2019},
	pages = {210},
}

@article{pedretti_tree-based_2021,
	title = {Tree-based machine learning performed in-memory with memristive analog {CAM}},
	volume = {12},
	issn = {2041-1723},
	doi = {10.1038/s41467-021-25873-0},
	language = {en},
	number = {1},
	urldate = {2026-02-17},
	journal = {Nature Communications},
	author = {Pedretti, Giacomo and Graves, Catherine E. and Serebryakov, Sergey and Mao, Ruibin and Sheng, Xia and Foltin, Martin and Li, Can and Strachan, John Paul},
	month = oct,
	year = {2021},
	pages = {5806},
}

@article{de_la_cruz_explainable_2024,
	title = {Explainable hypoglycemia prediction models through dynamic structured grammatical evolution},
	volume = {14},
	issn = {2045-2322},
	doi = {10.1038/s41598-024-63187-5},
	language = {en},
	number = {1},
	urldate = {2026-02-17},
	journal = {Scientific Reports},
	author = {De La Cruz, Marina and Garnica, Oscar and Cervigon, Carlos and Velasco, Jose Manuel and Hidalgo, J. Ignacio},
	month = jun,
	year = {2024},
	pages = {12591}
}

@article{gadaleta_prediction_2019,
	title = {Prediction of {Adverse} {Glycemic} {Events} {From} {Continuous} {Glucose} {Monitoring} {Signal}},
	volume = {23},
	copyright = {https://ieeexplore.ieee.org/Xplorehelp/downloads/license-information/IEEE.html},
	issn = {2168-2194, 2168-2208},
	doi = {10.1109/JBHI.2018.2823763},
	number = {2},
	urldate = {2026-02-17},
	journal = {IEEE Journal of Biomedical and Health Informatics},
	author = {Gadaleta, Matteo and Facchinetti, Andrea and Grisan, Enrico and Rossi, Michele},
	month = mar,
	year = {2019},
	pages = {650--659},
}

@article{zhu_personalized_2023,
	title = {Personalized {Blood} {Glucose} {Prediction} for {Type} 1 {Diabetes} {Using} {Evidential} {Deep} {Learning} and {Meta}-{Learning}},
	volume = {70},
	copyright = {https://ieeexplore.ieee.org/Xplorehelp/downloads/license-information/IEEE.html},
	issn = {0018-9294, 1558-2531},
	doi = {10.1109/TBME.2022.3187703},
	number = {1},
	urldate = {2026-02-17},
	journal = {IEEE Transactions on Biomedical Engineering},
	author = {Zhu, Taiyu and Li, Kezhi and Herrero, Pau and Georgiou, Pantelis},
	month = jan,
	year = {2023},
	pages = {193--204},
}

@inproceedings{yang_joint_2022,
	address = {Singapore, Singapore},
	title = {Joint {Hypoglycemia} {Prediction} and {Glucose} {Forecasting} via {Deep} {Multi}-{Task} {Learning}},
	copyright = {https://doi.org/10.15223/policy-029},
	isbn = {978-1-6654-0540-9},
	doi = {10.1109/ICASSP43922.2022.9746129},
	urldate = {2026-02-17},
	booktitle = {{ICASSP} 2022 - 2022 {IEEE} {International} {Conference} on {Acoustics}, {Speech} and {Signal} {Processing} ({ICASSP})},
	publisher = {IEEE},
	author = {Yang, Mu and Dave, Darpit and Erraguntla, Madhav and Cote, Gerard L. and Gutierrez-Osuna, Ricardo},
	month = may,
	year = {2022},
	pages = {1136--1140},
}

@article{marling_ohiot1dm_2020,
	title = {The {OhioT1DM} {Dataset} for {Blood} {Glucose} {Level} {Prediction}: {Update} 2020},
	volume = {2675},
	issn = {1613-0073},
	shorttitle = {The {OhioT1DM} {Dataset} for {Blood} {Glucose} {Level} {Prediction}},
	language = {eng},
	journal = {CEUR workshop proceedings.},
	author = {Marling, Cindy and Bunescu, Razvan},
	month = sep,
	year = {2020},
	pages = {71--74},
}

@inproceedings{timofte_sparse_2011,
	address = {Dundee},
	title = {Sparse {Representation} {Based} {Projections}},
	isbn = {978-1-901725-43-8},
	doi = {10.5244/C.25.61},
	language = {en},
	urldate = {2026-02-17},
	booktitle = {Procedings of the {British} {Machine} {Vision} {Conference} 2011},
	publisher = {British Machine Vision Association},
	author = {Timofte, Radu and Gool, Luc Van},
	year = {2011},
	pages = {61.1--61.12},
}

@article{kang_194-njdecision_2018,
	title = {A 19.4-{nJ}/{Decision}, 364-{K} {Decisions}/s, {In}-{Memory} {Random} {Forest} {Multi}-{Class} {Inference} {Accelerator}},
	volume = {53},
	copyright = {https://ieeexplore.ieee.org/Xplorehelp/downloads/license-information/IEEE.html},
	issn = {0018-9200, 1558-173X},
	doi = {10.1109/JSSC.2018.2822703},
	number = {7},
	urldate = {2026-02-17},
	journal = {IEEE Journal of Solid-State Circuits},
	author = {Kang, Mingu and Gonugondla, Sujan K. and Lim, Sungmin and Shanbhag, Naresh R.},
	month = jul,
	year = {2018},
	pages = {2126--2135},
}

@article{shin_neuraltree_2022,
	title = {{NeuralTree}: {A} 256-{Channel} 0.227-mu{J}/{Class} {Versatile} {Neural} {Activity} {Classification} and {Closed}-{Loop} {Neuromodulation} {SoC}},
	volume = {57},
	copyright = {https://ieeexplore.ieee.org/Xplorehelp/downloads/license-information/IEEE.html},
	issn = {0018-9200, 1558-173X},
	shorttitle = {{NeuralTree}},
	doi = {10.1109/JSSC.2022.3204508},
	number = {11},
	urldate = {2026-02-17},
	journal = {IEEE Journal of Solid-State Circuits},
	author = {Shin, Uisub and Ding, Cong and Zhu, Bingzhao and Vyza, Yashwanth and Trouillet, Alix and Revol, Emilie C. M. and Lacour, Stephanie P. and Shoaran, Mahsa},
	month = nov,
	year = {2022},
	pages = {3243--3257}
}

@misc{r_a_fisher_iris_1936,
	title = {Iris},
	url = {https://archive.ics.uci.edu/dataset/53},
	doi = {10.24432/C56C76},
	urldate = {2026-02-17},
	publisher = {UCI Machine Learning Repository},
	author = {{R. A. Fisher}},
	year = {1936},
}

@misc{william_wolberg_breast_1993,
	title = {Breast {Cancer} {Wisconsin} ({Diagnostic})},
	url = {https://archive.ics.uci.edu/dataset/17},
	doi = {10.24432/C5DW2B},
	urldate = {2026-02-17},
	publisher = {UCI Machine Learning Repository},
	author = {William Wolberg, Olvi Mangasarian},
	year = {1993},
}

@article{lee_vocabulary_2015,
  author={Lee, Kyuho Jason and Kim, Gyeonghoon and Park, Junyoung and Yoo, Hoi-Jun},
  journal={IEEE Journal of Solid-State Circuits}, 
  title={A Vocabulary Forest Object Matching Processor With 2.07 M-Vector/s Throughput and 13.3 nJ/Vector Per-Vector Energy for Full-HD 60 fps Video Object Recognition}, 
  year={2015},
  volume={50},
  number={4},
  pages={1059-1069},
  doi={10.1109/JSSC.2014.2380790}
}

@ARTICLE{shih_dr_2023,
  author={Shih, Xin-Yu and Chiu, Yao and Wu, Hsiang-En},
  journal={IEEE Transactions on Circuits and Systems I: Regular Papers}, 
  title={Design and Implementation of Decision-Tree (DT) Online Training Hardware Using Divider-Free GI Calculation and Speeding-Up Double-Root Classifier}, 
  year={2023},
  volume={70},
  number={2},
  pages={759-771},
  doi={10.1109/TCSI.2022.3222515}
}

@INPROCEEDINGS{liu_bnn_2025,
  author={Liu, Jianbo and Enciso, Zephan and Cheng, Boyang and Pei, Likai and Davis, Steven and Qin, Yifan and Jia, Zhenge and Hu, Xiaobo Sharon and Shi, Yiyu and Cao, Ningyuan},
  booktitle={2025 IEEE International Solid-State Circuits Conference (ISSCC)}, 
  title={15.3 A 65nm Uncertainty-Quantifiable Ventricular Arrhythmia Detection Engine with $\mathbf{1.75}\boldsymbol{\mu}\mathbf{J}$ Per Inference}, 
  year={2025},
  volume={68},
  number={},
  pages={1-3},
  doi={10.1109/ISSCC49661.2025.10904610}
}

@INPROCEEDINGS{bhat_gradient_2022,
  author={Bhat, Ashwin and Assoa, Adou Sangbone and Raychowdhury, Arijit},
  booktitle={2022 IFIP/IEEE 30th International Conference on Very Large Scale Integration (VLSI-SoC)}, 
  title={Gradient Backpropagation based Feature Attribution to Enable Explainable-AI on the Edge}, 
  year={2022},
  volume={},
  number={},
  pages={1-6},
  doi={10.1109/VLSI-SoC54400.2022.9939601}
}

@INPROCEEDINGS{cheng_neuromorphi_2024,
  author={Cheng, Boyang and Liu, Jianbo and Davis, Steven and Enciso, Zephan M. and Pei, Likai and Chang, Muya and Cao, Ningyuan},
  booktitle={2024 IEEE Symposium on VLSI Technology and Circuits (VLSI Technology and Circuits)}, 
  title={A 65nm Neuromorphic Bio-Signal Encoder with Compute-in-Entropy Architecture 7.13nJ Privacy-Preserving Encoding and 2.38Mb/mm2 Item Memory Density}, 
  year={2024},
  volume={},
  number={},
  pages={1-2},
  doi={10.1109/VLSITechnologyandCir46783.2024.10631333}
}

@ARTICLE{enciso_bnn_2026,
  author={Enciso, Zephan M. and Liu, Jianbo and Cheng, Boyang and Pei, Likai and Davis, Steven and Qin, Yifan and Jia, Zhenge and Hu, Xiaobo Sharon and Shi, Yiyu and Niemier, Michael and Cao, Ningyuan},
  journal={IEEE Journal of Solid-State Circuits}, 
  title={A 350-pW Implantable Ventricular Arrhythmia Detection Engine With Bayesian Uncertainty Quantification in 65-nm CMOS}, 
  year={2026},
  volume={},
  number={},
  pages={1-11},
  doi={10.1109/JSSC.2026.3669040}
}

@misc{frosst_soft_2017,
      title={Distilling a Neural Network Into a Soft Decision Tree}, 
      author={Nicholas Frosst and Geoffrey Hinton},
      year={2017},
      eprint={1711.09784},
      archivePrefix={arXiv},
      primaryClass={cs.LG},
      url={https://arxiv.org/abs/1711.09784}, 
}

@ARTICLE{xie_benchmarking_2020,
  author={Xie, Jinyu and Wang, Qian},
  journal={IEEE Transactions on Biomedical Engineering}, 
  title={Benchmarking Machine Learning Algorithms on Blood Glucose Prediction for Type I Diabetes in Comparison With Classical Time-Series Models}, 
  year={2020},
  volume={67},
  number={11},
  pages={3101-3124},
  doi={10.1109/TBME.2020.2975959}
}

@article{ada_glycemic_2023,
    author = {American Diabetes Association Professional Practice Committee},
    title = {Glycemic Goals and Hypoglycemia: Standards of Care in Diabetes—2024},
    journal = {Diabetes Care},
    volume = {47},
    number = {Supplement-1},
    pages = {S111-S125},
    year = {2023},
    month = {12},
    doi = {10.2337/dc24-S006}
}

@inproceedings{cai_vibnn_2018,
author = {Cai, Ruizhe and Ren, Ao and Liu, Ning and Ding, Caiwen and Wang, Luhao and Qian, Xuehai and Pedram, Massoud and Wang, Yanzhi},
title = {VIBNN: Hardware Acceleration of Bayesian Neural Networks},
year = {2018},
isbn = {9781450349116},
publisher = {Association for Computing Machinery},
address = {New York, NY, USA},
url = {https://doi.org/10.1145/3173162.3173212},
doi = {10.1145/3173162.3173212},
booktitle = {Proceedings of the Twenty-Third International Conference on Architectural Support for Programming Languages and Operating Systems},
pages = {476–488},
numpages = {13},
location = {Williamsburg, VA, USA},
series = {ASPLOS '18}
}

@ARTICLE{liu_neuro_symblic_2026,
  author={Liu, Che-Kai and Wan, Zishen and Noh, Young-Seok and Ibrahim, Mohamed and Spetalnick, Samuel D. and Krishna, Tushar and Khwa, Win-San and Sanjay Lele, Ashwin and Chih, Yu-Der and Chang, Meng-Fan and Raychowdhury, Arijit},
  journal={IEEE Journal of Solid-State Circuits}, 
  title={A 40-nm Programmable Heterogeneous SoC With 5.625/0.85 MB RRAM/SRAM for Accelerating Neuro-Symbolic AI Models}, 
  year={2026},
  volume={},
  number={},
  pages={1-16},
  doi={10.1109/JSSC.2026.3658288}}
 
% \vspace{11pt}

% \bf{If you include a photo:}\vspace{-33pt}
% \begin{IEEEbiography}[{\includegraphics[width=1in,height=1.25in,clip,keepaspectratio]{fig1}}]{Michael Shell}
% Use $\backslash${\tt{begin\{IEEEbiography\}}} and then for the 1st argument use $\backslash${\tt{includegraphics}} to declare and link the author photo.
% Use the author name as the 3rd argument followed by the biography text.
% \end{IEEEbiography}

% \vspace{11pt}

% \bf{If you will not include a photo:}\vspace{-33pt}
% \begin{IEEEbiographynophoto}{John Doe}
% Use $\backslash${\tt{begin\{IEEEbiographynophoto\}}} and the author name as the argument followed by the biography text.
% \end{IEEEbiographynophoto}

\vfill

\end{document}